\documentclass[aps,prd,preprint,a4paper,showpacs]{revtex4}
\usepackage{graphicx,natbib}
\usepackage{amsmath}
\newcommand{\dis}{\displaystyle}
\newcommand{\bi}[1]{\mbox{\boldmath ${#1}$}}
\begin{document}
\title{Flavor Mass and Mixing and $S_3$ Symmetry \\
-- An $S_3$ Invariant Model Reasonable to All --}
\vspace*{1cm}
\author{T. Teshima}
\email{teshima@isc.chubu.ac.jp}
\affiliation{Department of Applied Physics,  Chubu University, 
Kasugai 487-8501, Japan}
\begin{abstract}
We assume that weak bases of flavors $(u, c)_{L,R}$, $(d,s)_{L,R}$, $(e, \mu)
_{L,R}$, $(\nu_e, \nu_\mu)_{L,R}$ are the $S_3$ doublet and $t_{L,R}$, 
$b_{L,R}$, $\tau_{L,R}$, ${\nu_\tau}_{L,R}$ are the $S_3$ singlet and further 
there are $S_3$ doublet Higgs $(H_D^1, H_D^2)$ and $S_3$ singlet Higgs $H_S$. 
We suggest an $S_3$ invariant Yukawa interaction, in which masses 
caused from the interaction of $S_3$ singlet flavors and Higgs is very large and 
masses caused from interactions of $S_3$ doublet flavors and Higgs are very small, 
and the vacuum expectation value $\langle|H_D^1|\rangle_{0}$ is rather small 
compared to the $\langle|H_D^2|\rangle_{0}$. 
In this model, we can explain the quark sector mass hierarchy, flavor mixing 
$V_{\rm CKM}$ and measure of CP violation naturally. 
The leptonic sector mass hierarchy and flavor mixing described by $V_{\rm MNS}$ 
having one-maximal and one-large mixing character can also be explained 
naturally with no other symmetry restriction.
In our model, an origin of Cabibbo angle is the ratio $\lambda=\langle|H_D^1|
\rangle_{0}/\langle|H_D^2|\rangle_{0}$ and an origin of CP violation is the 
phase of $H_D^1$.
\end{abstract}
\preprint{CU-TP/05-09}
\maketitle
\section{Introduction}
Many authors have succeeded in explaining the quark mass hierarchy and quark mixing 
characterized by $V_{\rm CKM}$ and neutrino mixing $V_{\rm MNS}$ having the large 
mixing character confirmed by recent experiments \cite{ATMOS,SOLAR}, using 
democratic/universal model \cite{HARARI,KOIDE1,TANIMOTO1,KOIDE2,GATTO, BRANCO1,
TESHIMA1,TESHIMA2,TESHIMA3,BRANCO2,BRANVCO3} or Fritzsch-type model 
\cite{FRITZSCH,FUKUGITA1,FUKUGITA2}. 
However, many works adopting the democratic/universal model assume small mass terms 
deviated from  democratic/universal masses as to be any violations to a symmetry 
reflecting in democratic/universal character. 
In Fritzsch-type model, the principle why Fritzsch-type mass matrix is 
assumed is not clear. 
\par
Many authors \cite{PAKUBASA,SUGAWARA,KUBO} have considered the $S_3$ symmetry as 
a symmetry reflecting in democratic/universal character. 
Especially, Kubo {\it et. al.} \cite{KUBO} suggested an outstanding model 
considering the small mass terms deviated from democratic/universal masses as to 
be $S_3$ invariant. 
However, their model assumes an additional symmetry to explain the neutrino 
mixing having bi-maximal mixing character. 
\par 
In this work, we suggest a model in which the small mass terms deviated from 
democratic/universal mass are constructed by an $S_3$ invariant manner, and 
the quark mass hierarchy and mixing and neutrino mixing having one-maximal and 
one-large mixing character confirmed by recent experiments \cite{ATMOS,SOLAR} is 
explained without any other symmetry restriction. 
\section{$S_3$ invariant model}
In the democratic/universal models, the mass matrix for the quark and lepton fields 
is assumed to be composed of the completely universal term with respect to an basis 
$\displaystyle{{\bi v}_{L,R}=^t(v_1, v_2, v_3)}_{L,R}$ ($t$ denotes the transpose 
of matrix or vector) and the small violation term or small gap from the universal 
term as following form \cite{HARARI,KOIDE1,TANIMOTO1,KOIDE2,GATTO,BRANCO1,TESHIMA1,
TESHIMA2,TESHIMA3,BRANCO2,BRANVCO3}, 
\begin{equation}
M=\Gamma \left[
               \left(\begin{array}{ccc}
                1&1&1\\1&1&1\\1&1&1
                \end{array}\right)-
                \left(\begin{array}{ccc}
                \delta_4&\delta_1&\delta_2\\
                \delta_1&\delta_5&\delta_3\\
                \delta_2&\delta_3&0
                \end{array}\right)
          \right],\ \ \ 
               \delta_i\ll1\ \  (i=1,2,3,4,5),
\end{equation}
where we assumed a symmetric mass matrix, because it is natural to assume that 
there is no difference in kinematical quantity producing mass between $u_L$ 
and $u_R$, then the magnitudes of the Yukawa interaction of, e.g., 
$u_Ls_R$ and $s_Lu_R$ are same. 
In next section, we consider the CP violation, then we will treat an Hermitian 
mass matrix there. 
In the Fritzsch-type models\cite{FRITZSCH,FUKUGITA1,FUKUGITA2}, the mass matrix 
has the following form,  
\begin{equation}
M=
  \left(\begin{array}{ccc}
  0&A&0\\A&0&B\\0&B&C
  \end{array}\right),\ \ \ A \ll B \ll C.
\end{equation}
\par
Mass matrix (1) is transformed by an orthogonal matrix $T$ with three small angles  
to the form, in which basis vectors are transformed a little, 
\begin{equation}
\begin{array}{l}
        {\bi v}'_{L,R}=^t\!(v'_1,v'_2,v'_3)_{L,R}={T^{-1}}^t(v_1,v_2,v_3)_{L,R},\\
        M'=T^{-1}MT=\Gamma' \left[
               \left(\begin{array}{ccc}
                1&1&1\\1&1&1\\1&1&1
                \end{array}\right)-
                \left(\begin{array}{ccc}
                0&\delta_1'&\delta_2'\\
                \delta_1'&0&\delta_3'\\
                \delta_2'&\delta_3'&0
                \end{array}\right)
          \right],\ \ \ 
               \delta_i'\ll1\ \  (i=1,2,3),
\end{array}               
\end{equation}
or
\begin{equation}
\begin{array}{l}
        {\bi v}''_{L,R}=^t\!(v''_1,v''_2,v''_3)_{L,R}={T^{-1}}^t\!(v_1,v_2,v_3)_
        {L,R},\\
M''=T^{-1}MT=\Gamma'' \left[
               \left(\begin{array}{ccc}
                1&1&1\\1&1&1\\1&1&1
                \end{array}\right)-
                \left(\begin{array}{ccc}
                \delta_4''&0&0\\
                0&\delta_5''&0\\
                0&0&\delta_6''
                \end{array}\right)
          \right].\ \ \ 
               \delta_i''\ll1\ \  (i=4,5,6).
\end{array}                  
\end{equation}
Thus, the basis vector for flavor fields depends on the mass matrix pattern assumed 
in the democratic/universal model.
This fact suggests that one cannot define a basis vector definitely in the 
democratic/universal model. 
We will discuss on this problem later in this section. 
\par
Hereafter, we assume the mass matrix pattern (4) for we used this mass matrix in 
our early works \cite{TESHIMA1,TESHIMA2,TESHIMA3}. 
We define the eigen vectors ${\bi f}_{L,R}=^t\!(f_1,f_2,f_3)_{L,R}$ corresponding 
to the eigenvalues $(0,\ 0\ 3\Gamma)$ for the first term in (4) (we drop the 
prime (') in (4) hereafter), 
\begin{equation}
\begin{array}{l}
  \left( \begin{array}{c}f_1\\f_2\\f_3\end{array}\right)_{L,R}
  = \left( \begin{array}{c}
              \frac1{\sqrt{2}}v_1-\frac1{\sqrt{2}}v_2\\
              \frac1{\sqrt{6}}v_1+\frac1{\sqrt{6}}v_2-\frac2{\sqrt{6}}v_3\\
              \frac1{\sqrt{3}}v_1+\frac1{\sqrt{3}}v_2+\frac1{\sqrt{3}}v_3
               \end{array}\right)_{L,R}
  =\left( \begin{array}{ccc}
              \frac1{\sqrt{2}}&-\frac1{\sqrt{2}}&0\\
              \frac1{\sqrt{6}}&\frac1{\sqrt{6}}&-\frac2{\sqrt{6}}\\
              \frac1{\sqrt{3}}&\frac1{\sqrt{3}}&\frac1{\sqrt{3}}
               \end{array}\right)
    \left( \begin{array}{c}v_1\\v_2\\v_3\end{array}\right)_{L,R}\\
  \ \ \ \ \ \ \ \ \ \ \ ={T_0^{-1}}^t(p_1, p_2, p_3)_{L,R},\\
  T_0= \left( \begin{array}{ccc}
              \frac1{\sqrt{2}}&\frac1{\sqrt{6}}&\frac1{\sqrt{3}}\\
              -\frac1{\sqrt{2}}&\frac1{\sqrt{6}}&\frac1{\sqrt{3}}\\
              0&-\frac2{\sqrt{6}}&\frac1{\sqrt{3}}
               \end{array}\right),             
\end{array}
\end{equation}
where the definition of $T_0$ is the inverse of the $T_0$ used in our previous 
papers \cite{TESHIMA1,TESHIMA2,TESHIMA3}. 
In this basis vector ${\bi f}_{L,R}=^t\!(f_1,f_2,f_3)_{L,R}$, the mass matrix (4) 
is expressed as 
\begin{equation}
M_f=T^{-1}_0MT_0=\Gamma\left(
    \begin{array}{ccc}
    \delta_1&\frac1{\sqrt{3}}(\delta_2-\delta_3)&-\frac1{\sqrt{6}}
               (\delta_2-\delta_3)\\
    \frac1{\sqrt{3}}(\delta_2-\delta_3)&\frac13(-\delta_1+2\delta_2+2\delta_3)&
                \frac1{3\sqrt{2}}(-2\delta_1+\delta_2+\delta_3)\\
    -\frac1{\sqrt{6}}(\delta_2-\delta_3)&\frac1{3\sqrt{2}}(-2\delta_1+\delta_2+
                \delta_3)&\frac13(9-2\delta_1-2\delta_2-2\delta_3)
    \end{array}\right).            
\end{equation}
Diagonalising this mass matrix $M_f$ by the transformation matrix $U(\delta_1,
\delta_2,\delta_3)$ nearly equal to identity, the mass eigenvalues $m_1, m_2, m_3$ 
and eigenvectors corresponding to these masses are 
obtained as 
\begin{equation}
\begin{array}{l}
{\rm diag}(m_1, m_2, m_3)=U^{-1}(\delta_1,\delta_2,\delta_3)M_fU(\delta_1,
\delta_2,\delta_3)\\
\  \ m_1\approx \left[\frac13(\delta_1+\delta_2+\delta_3)-\frac13
      \xi\right]\Gamma\approx \delta_1\Gamma,\\
\  \ m_2\approx \left[\frac13(\delta_1+\delta_2+\delta_3)+\frac13
      \xi\right]\Gamma\approx\frac23(\delta_2+\delta_3)\Gamma,\\
\  \ m_3\approx \left[3-\frac23(\delta_1+\delta_2+\delta_3)
      \right]\Gamma\approx3\Gamma,\\
\ \ \ \ \ \ \xi=\left[(\delta_2+\delta_3-2\delta_1)^2+3(\delta_2-\delta_3)^2
      \right]^{1/2}\\
^t(f^m_1, f^m_2, f^m_3)_{L,R}=U^{-1}(\delta_1,\delta_2,\delta_3)^t
(f_1, f_2, f_3)_{L,R},
\end{array}      
\end{equation}
where $(f^m_1, f^m_2, f^m_3)_{L,R}$ is mass basis. 
In our paper \cite{TESHIMA1}, we analyzed the quark sector mass 
hierarchy and quark mixing matrix $V_{\rm CKM}$ numerically using this mass matrix 
(4), and get the numerical results that 
\begin{equation}
\delta_1,\ \delta_3-\delta_2\ll\delta_2, \delta_3\ll1.
\end{equation} 
The basis $^t(f_1, f_2, f_3)_{L,R}$ may be considered as a weak basis 
because the mass matrix (6) in the basis $^t(f_1, f_2, f_3)_{L,R}$ has non 
zero nondiagonal elements and then produce the flavor mixing matrix not equal to 
identity. 
The Fritzsch-type mass matrix (2) is also considered to be a model based on the 
weak basis because of the parameter hierarchy $A\ll B\ll C$. 
It should be noted that the weak basis based on above democratic/universal-type 
model and Fritzsch-type model has an ambiguity depending on which type mass 
pattern is assumed. 
\par
The basic idea of the democratic/universal-type model is as follows;
the Yukawa interaction strengths composed of the basis field $(v_1, v_2, 
v_3)_{L,R}$ are almost same values, then  the main part of the mass matrix is 
invariant under the permutation of the basis fields $(v_1, v_2, v_3)_{L,R}$, i.e., 
the basis $(v_1, v_2, v_3)_{L,R}$ is considered as the basis field of the 
permutation group $S_3$ and the main part of the mass matrix is the product of 
the $S_3$ singlets $\displaystyle{f_S^{_{L,R}}=(v_1+v_2+v_3)_{L,R}/\sqrt{3}}$. 
Mass matrix (4) has the small parameters $\delta_1, \delta_2, \delta_3$ in 
addition to this main mass term. 
We assume that these small parameters are produced by the way in which 
$S_3$ invariance is satisfied. 
The basis $(v_1, v_2, v_3)_{L,R}$ are translated to $S_3$ singlet 
$f_S^{_{L,R}}$ and doublet $f_D^{_{L,R}}$ as 
\begin{equation}
\begin{array}{l}
S_3\ {\rm singlet}\ \ \ f_S^{_{L,R}}=\frac1{\sqrt{3}}(v_1+v_2+v_3)_
{L,R},\\
S_3\ {\rm doublet}\ \ \ {\bi f}_D^{_{L,R}}=\left(\begin{array}{c}f_1^{_{L,R}}\\
f_2^{_{L,R}}\end{array}\right)
=\left(\begin{array}{c}\frac1{\sqrt{2}}(v_1-v_2)_{L,R}\\
     \frac1{\sqrt{6}}(v_1+v_2-2v_3)_{L,R}\end{array}\right).
\end{array}
\end{equation}
By using these $S_3$ singlet and doublet of flavors, we can make the following 
$S_3$ invariant Yukawa interaction;
$$
\Gamma_S\bar{f^L_S}f^R_SH+\Gamma_D\bar{{\bi f}^L_D}{\bi f}^R_DH+h.c., 
$$
where $H$ is the neutral part of $SU(2)_L$ Higgs doublet field and is considered 
as $S_3$ singlet. 
The mass matrix of this Yukawa interaction is written as 
$$
\left(\begin{array}{ccc}
  \Gamma_Dh&0&0\\0&\Gamma_Dh&0\\0&0&\Gamma_Sh
  \end{array}\right),
$$  
where $h=\langle H \rangle_0$ is the vacuum expectation value of $H$. 
One can get the mass of doublet $f_1$ and $f_2$, but cannot get the mass 
hierarchy $m_1\ll m_2
\ll m_3 $. 
Thus, we introduce an $S_3$ doublet of the Higgs field 
\begin{equation}
S_3\ {\rm singlet}\ :\ H_S,\ \ S_3\ {\rm doublet}\ :\ {\bi H}_D=^t\!(H_D^1, H_D^2), 
\end{equation} 
and make the $S_3$ invariant Yukawa interactions including the neutral $S_3$ 
doublet ${\bi H}_D$ \cite{KUBO};
\begin{eqnarray}
&-{\cal L_D}=\Gamma_S\bar{f^L_S}f^R_SH_S+\Gamma_D^1{\bar{\bi f}^L_D}{\bi f}^R_DH_S+
\Gamma_D^2[(\bar{f_1^L}f_2^R+\bar{f_2^L}f_1^R)H_D^1+
(\bar{f_1^L}f_1^R-\bar{f_2^L}f_2^R)H_D^2]\nonumber\\
&+\Gamma_D^3(\bar{\bi f}_D^L{\bi H}_Df_S^R+\bar{f_S^L}^t\!{\bi H}_D
{\bi f}_D^R)+h.c.,  
\end{eqnarray}
where we used the fact that the $S_3$ doublet can be made from the tensor product 
of $\bar{\bi f}_D^L$ and ${\bi f}_D^R$ \cite{PAKUBASA,SUGAWARA,KUBO} as 
$$
\left(\begin{array}{c}
    \bar{f_1^L}f_2^R+\bar{f_2^L}f_1^R\\
    \bar{f_1^L}f_1^R-\bar{f_2^L}f_2^R
\end{array}\right).   
$$
In general $H_S$ and ${\bi H_D}$ are complex fields, but in this section we assume  
that these fields are real for simplicity of discussion. 
In next section discussing the quark sector mass and mixing, we consider the case 
that $H_S$ and ${\bi H_D}$ are complex fields for we discuss the CP violation. 
The mass matrix corresponding to Yukawa interaction (11) is expressed as 
\begin{eqnarray}
&&M_f=\left(\begin{array}{ccc}
     \Gamma_D^1h+\Gamma_D^2h_2&\Gamma_D^2h_1&\Gamma_D^3h_1\\
     \Gamma_D^2h_1&\Gamma_D^1h-\Gamma_D^2h_2&\Gamma_D^3h_2\\
     \Gamma_D^3h_1&\Gamma_D^3h_2&\Gamma_Sh
     \end{array}\right),\ \ \ \ \ \ \\
&&\ \ h=\langle H_S\rangle_0,\  h_1=\langle H_D^1\rangle_0,\  h_2=\langle H_D^2
\rangle_0.\nonumber
\end{eqnarray}
Because $\Gamma_D$ is considered to be very small compared to $\Gamma_S$, this 
mass matrix can be similar to the mass matrix (6), if the following condition 
is satisfied,
\begin{equation}
\Gamma_D^1h+\Gamma_D^2h_2 \ll \Gamma_D^1h-\Gamma_D^2h_2 \approx \Gamma_D^3h_2 \ll 
\Gamma_Sh, \ {h_1}\ll{h_2}.
\end{equation}
Here, it should be stressed that the expectation value of $H_D^1$ has to be 
rather small compared to that of $H_D^2$ so as to produce the realistic quark 
mass hierarchy and mixing matrix $V_{\rm CKM}$.
\par
Next we consider the neutrino mass. For neutrino, we assume that there are very 
large Majorana masses constructed from the right-handed neutrinos and very small 
neutrino masses are obtained from the see-saw mechanism \cite{SEESAW}. 
The neutrino Dirac mass matrix is obtained from the Yukawa interaction similar to 
Eq.~(11). 
The Higgs fields are necessary in the Yukawa interaction for Dirac mass because 
$SU(2)_L$ doublet left-handed Dirac field could not make $SU(2)_L$ invariant 
interactions without $SU(2)_L$ doublet of Higgs fields.
But there is no reason that Higgs field is necessary for the Yukawa interaction 
producing Majorana mass. 
Thus we assume that the Majorana mass is obtained from the $S_3$ invariant Yukawa 
interaction containing only right handed neutrino ${\bi \nu}_D^R=^t\!(\nu_1^R, 
\nu_2^R),\ \nu_S^R$ and no Higgs field \cite{KUBO} as 
\begin{equation}
-{\cal L_M}=\Gamma_S^M{^t}\nu^R_SC^{-1}\nu^R_S+\Gamma_D^M{^t}{\bi \nu}^R_DC^{-1}
{\bi \nu}^R_D.
\end{equation}
If $\Gamma_D^M\ll\Gamma_S^M$ as case of the Dirac neutrino mass, we can explain 
the one-maximal and one-large neutrino mixing character in this model without 
any other symmetry restriction. 
We will discuss this problem in section 4 in 
detail.
\par 
Finally, we comment on the weak basis, which is the flavor basis in weak interaction. In $SU(3)_C\times SU(2)_L\times U(1)_Y$ standard model, there is no definite 
definition determining the weak basis because we have no definite criterion for 
assuming a mass matrix pattern and then Yukawa interaction constructing the flavor 
masses. 
However, if one assume the $S_3$ symmetry for flavor and define the weak basis 
using the $S_3$ singlet and doublet defined in Eq. (9) as follows;
\begin{equation}
\left(\begin{array}{c}f_1^{L,R}\\f_2^{L,R}\\
f_S^{L,R}\end{array}\right)=
\left(\begin{array}{c}u^{L,R}\\c^{L,R}\\
t^{L,R}\end{array}\right),
\left(\begin{array}{c}d^{L,R}\\s^{L,R}\\
b^{L,R}\end{array}\right),
\left(\begin{array}{c}e^{L,R}\\\mu^{L,R}\\
\tau^{L,R}\end{array}\right),
\left(\begin{array}{c}{\nu_e}^{L,R}\\
{\nu_\mu}^{L,R}\\{\nu_\tau}^{L,R}\end{array}\right),
\end{equation}
the $S_3$ invariant Yukawa interaction is determined uniquely as Eqs. (11) and 
(14), then the weak basis defined Eq. (15) is unique. 
\section{Quark mass and mixing}   
In this section, we consider the case that the neutral Higgs fields $H_S$ and 
${\bi H}_D$ are complex and then have phases. 
We assume the Yukawa interaction describing the masses of the $d$- and $u$-quark 
sector similar to Eq. (11);
\begin{eqnarray}
&-{\cal L_D}^d=\Gamma_S^d\bar{f^L_S}f^R_SH_S+\Gamma_D^{1d}{\bar{\bi f}^L_D}
{\bi f}^R_DH_S+\Gamma_D^{2d}[(\bar{f_1^L}f_2^RH_D^{1}+\bar{f_2^L}f_1^RH_D^{1*})+
(\bar{f_1^L}f_1^R-\bar{f_2^L}f_2^R)H_D^2]\nonumber\\
&+\Gamma_D^{3d}[(\bar{f}_1^LH_D^1+\bar{f}_2^LH_D^2)f_S^R+\bar{f_S^L}({H}^{1*}_D
f_1^R+H_D^2f_2^R)]+h.c.,  {\rm \ for\ \mbox{$d$-}quark\ sector}\nonumber\\
&-{\cal L_D}^u=\Gamma_S^u\bar{f^L_S}f^R_SH_S^*+\Gamma_D^{1u}{\bar{\bi f}^L_D}
{\bi f}^R_DH_S^*+\Gamma_D^{2u}[(\bar{f_1^L}f_2^RH_D^{1*}+\bar{f_2^L}f_1^RH_D^{1})+
(\bar{f_1^L}f_1^R-\bar{f_2^L}f_2^R)H_D^{2*}]\nonumber\\
&+\Gamma_D^{3u}[(\bar{f}_1^LH_D^{1*}+\bar{f}_2^LH_D^{2*})f_S^R+
\bar{f}_S^L({H}_D^1f_1^R+H_D^{2*}f_2^R)]+h.c.,  {\rm \ for\ \mbox{$u$-}
quark\ sector}\\
&\left(\begin{array}{c}{\bi f}_D^{L,R}\\
f_S^{L,R}\end{array}\right)=
\left(\begin{array}{c}f_1^{L,R}\\f_2^{L,R}\\
f_S^{L,R}\end{array}\right)=
\left(\begin{array}{c}u^{L,R}\\c^{L,R}\\
t^{L,R}\end{array}\right),
\left(\begin{array}{c}d^{L,R}\\s^{L,R}\\
b^{L,R}\end{array}\right), \ \ \ 
\left(\begin{array}{c}{\bi H}_D\\H_S\end{array}\right)=
\left(\begin{array}{c}H_D^1\\H_D^2\\H_S\end{array}\right).\nonumber
\end{eqnarray}
This Yukawa interaction produces an helmitian mass matrix, as shown in just below. 
\par
The phase of $H_S$ is cancel out from the phase transformation of $f_S^L$ 
in the first term and ${\bi f}_D^L$ in the second term caused by the $SU(2)_L$ 
gauge freedom and then we can take the $H_S$ to be real. 
Here we assume that the phases of $f_S^L$ and ${\bi f}_D^L$ after the $SU(2)_L$ 
gauge transformation are the same as that of $f_S^R$ and ${\bi f}_D^R$, 
respectively. 
Thus the $H_S$ can be real and the masses produced from the first and second term 
are real. 
The phase of $H_D^2$ in third term has to be the same to that of the $H_S$, 
because the term containing $H_D^2$ in this forth term represents the diagonal mass 
and has to be real and then the $SU(2)_L$ gauge freedom of $f_1^L$ and $f_2^L$ 
have to be cancelled out by the phase of $H_D^2$.
From the same procedure, terms $\bar{f_2^L}H_D^2f_S^R$ and $\bar{f_S^L}H_D^2f_2^R$ 
in the forth term can be real. 
However, $H_D^1$ does not have to be the same phase to that of $H_S$, then the 
terms $\bar{f}_1^Lf_2^RH_D^1$ and $\bar{f}_2^Lf_1^RH_D^{1*}$ in the third term and 
the terms $\bar{f}_1^LH_D^1f_S^R$ and $\bar{f}_S^LH_D^{1*}f_1^R$ in the forth term 
may be complex. 
Thus we get the hermitian mass matrices for $d$ and $u$ quark sector as 
follows; 
\begin{equation}
\begin{array}{l}
M_d=\left(\begin{array}{ccc}
     \mu_1^d+\mu_2^d&\lambda\mu_2^de^{i\phi}&\lambda\mu_3^d
     e^{i\phi}\\
     \lambda\mu_2^de^{-i\phi}&\mu_1^d-\mu_2^d&\mu_3^d\\
     \lambda\mu_3^de^{-i\phi}&\mu_3^d&\mu_0^d
     \end{array}\right),\\ 
M_u=\left(\begin{array}{ccc}
     \mu_1^u+\mu_2^u&\lambda\mu_2^ue^{-i\phi}&\lambda\mu_3^u
     e^{-i\phi}\\
     \lambda\mu_2^ue^{i\phi}&\mu_1^u-\mu_2^u&\mu_3^u\\
     \lambda\mu_3^ue^{i\phi}&\mu_3^u&\mu_0^u
     \end{array}\right),
\end{array}     
\end{equation}
where we use the following parameterization, 
\begin{equation}
\begin{array}{l}
\mu_0^{d,u}=\langle\Gamma_S^{d,u}H_S\rangle_{SS},\ \ \mu_1^{d,u}=\langle
\Gamma_D^{1d,u}H_S\rangle_{11, 22},\ \ \mu_2^{d,u}=\langle\Gamma_D^{2d,u}H_D^2
\rangle_{11, 22},\nonumber\\
\lambda\mu_2^{d,u}=\langle\Gamma_D^{2d,u} |H_1|\rangle_{12, 21},\ \ \mu_3^{d,u}=
\langle\Gamma_D^{3d,u}H_D^2\rangle_{2S, S2},\ \ \lambda\mu_3^{d,u}=\langle
\Gamma_D^{3d,u}|H_D^1|\rangle_{1S, S1},\nonumber\\
\lambda=\displaystyle{\frac{\langle|H_D^1|\rangle_{0}}{\langle
|H_2|\rangle_{0}}},\ \ 
\phi={\rm phase\ of\ } H_D^1.\ 
\end{array}
\end{equation}
\par
Diagonalising the mass matrix (17) by unitary matrices $U(\mu_i^d,\phi)$ and 
$U(\mu_i^u,\phi)$, the CKM quark mixing matrix $V_{CKM}$ is defined as  
\begin{eqnarray}
&&{\rm diag}[m_d, m_s, m_b]=U^{-1}(\mu_i^d,\phi)M_dU(\mu_i^d,\phi),
\ \ \ ^t(d^m, s^m, b^m)_{L, R}=U^{-1}(\mu_i^d,\phi)^t(d, s, b)_{L, R},
\nonumber\\
&&{\rm diag}[m_u, m_c, m_t]=U^{-1}(\mu_i^u,\phi)M_uU(\mu_i^u,\phi),\ \ \ 
^t(u^m, c^m, t^m)_{L, R}=U^{-1}(\mu_i^u,\phi)^t(u, c, t)
_{L, R},
\nonumber\\
&&\hspace{1cm}V_{CKM}=U^{\dagger}(\mu_i^u,\phi)U(\mu_i^d,\phi), 
\end{eqnarray}
where $(d^m, s^m, b^m)$ and $(u^m, c^m, t^m)$ are the mass eigen states for the 
weak basis  $(d, s, b)$ and $(u, c, t)$, respectively.
Here, we count the number of parameters appearing in present model for quark 
sector: 10\ =\ 8 mass parameters $\mu_i^d$, $\mu_i^u$, $+$ 1 ratio of Higgs 
$|H_D^1|$ and $|H_D^2|$, $+$ 1 phase parameters $\phi={\rm phase\ of\ }H_D^1$.
Observable parameter number is 10: 6 for quark masses and 3 mixing 
angles and 1 phase in CKM matrix.
\par
We represent approximate expressions analytically for eigenvalues of quark 
masses and diagonalising unitary matrix. 
For the $d$-quark sector, masses and the diagonalising matrix are expressed 
as   
\begin{eqnarray}
&&m_d\approx\mu_1^d-\frac{1+\lambda^2}{2}\frac{\mu_3^{d2}}{\mu_0}+(\mu_2^d+
\frac{1-\lambda^2}{2}\frac{\mu_3^{d2}}{\mu_0^d})\frac{1}{\cos^2\theta_d-
\sin^2\theta_d}\approx\mu_1^d+\mu_2^d-\frac{\lambda^2\mu_3^{d2}}{\mu_0^d},
\nonumber\\
&&m_s\approx\mu_1^d-\frac{1+\lambda^2}{2}\frac{\mu_3^{d2}}{\mu_0}-(\mu_2^d+
\frac{1-\lambda^2}{2}\frac{\mu_3^{d2}}{\mu_0^d})\frac{1}{\cos^2\theta_d-
\sin^2\theta_d}\approx\mu_1^d-\mu_2^d-\frac{\mu_3^{d2}}{\mu_0^d},\\
&&m_b\approx\mu_0^d,\nonumber\\
&&\tan2\theta_d=\frac{\lambda\mu_2^d}{\mu_2^d+\frac{1-\lambda^2}
{2}\frac{\mu_3^{{d2}}}{\mu_0^d}},\nonumber\\
&&U(\mu_i^d,\phi)\approx\left(\begin{array}{ccc}
\cos\theta_d&-\sin\theta_de^{i\phi}&\frac{\lambda\mu_3^d}{\mu_0^d}
e^{i\phi}\\
\sin\theta_de^{-i\phi}&\cos\theta_d&\frac{\mu_3^d}{\mu_0^d}\\
(-\lambda\cos\theta_d-\sin\theta_d)e^{-i\phi}\frac{\mu_3^d}{\mu_0^d}&
(\lambda\sin\theta_d-\cos\theta_d)\frac{\mu_3^d}{\mu_0^d}&1
\end{array}\right).
\end{eqnarray}
For $u$-quark sector, just similar expressions are obtained replacing 
suffix $d$ with suffix $u$, 
\begin{eqnarray}
&&m_u\approx\mu_1^u-\frac{1+\lambda^2}{2}\frac{\mu_3^{u2}}{\mu_0}+(\mu_2^u+
\frac{1-\lambda^2}{2}\frac{\mu_3^{u2}}{\mu_0^u})\frac{1}{\cos^2\theta_u-
\sin^2\theta_u}\approx\mu_1^u+\mu_2^u-\frac{\lambda^2\mu_3^{u2}}{\mu_0^u},
\nonumber\\
&&m_c\approx\mu_1^u-\frac{1+\lambda^2}{2}\frac{\mu_3^{u2}}{\mu_0}-(\mu_2^u+
\frac{1-\lambda^2}{2}\frac{\mu_3^{u2}}{\mu_0^u})\frac{1}{\cos^2\theta_u-
\sin^2\theta_u}\approx\mu_1^u-\mu_2^u-\frac{\mu_3^{u2}}{\mu_0^u},\\
&&m_t\approx\mu_0^u,\nonumber\\
&&\tan2\theta_u=\frac{\lambda\mu_2^u}{\mu_2^u+\frac{1-\lambda^2}
{2}\frac{\mu_3^{{u2}}}{\mu_0^u}},\nonumber\\
&&U(\mu_i^u,\phi)\approx\left(\begin{array}{ccc}
\cos\theta_u&-\sin\theta_ue^{-i\phi}&\frac{\lambda\mu_3^u}{\mu_0^u}
e^{-i\phi}\\
\sin\theta_ue^{i\phi}&\cos\theta_u&\frac{\mu_3^u}{\mu_0^u}\\
(-\lambda\cos\theta_u-\sin\theta_u)^{i\phi}\frac{\mu_3^u}{\mu_0^u}&
(\lambda\sin\theta_u-\cos\theta_u)\frac{\mu_3^u}{\mu_0^u}&1
\end{array}\right).
\end{eqnarray}
The CKM matrix is also written analytically using the expressions Eqs. (20) 
and (22) as 
\begin{eqnarray}
&&V_{\rm CKM}=U^{\dagger}(\mu^u_{i}, \phi)U(\mu^d_i, \phi)
\nonumber\\
&&\ \ \ \ \ \ \ \ \approx\left(\begin{array}{c}
\cos\theta_u\cos\theta_d+\sin\theta_u\sin\theta_de^{i2\phi}\\
-\sin\theta_u\cos\theta_de^{i\phi}+\cos\theta_u\sin\theta_de^{-i\phi}\\
(\lambda\cos\theta_de^{i\phi}+\sin\theta_d)\frac{\mu^u_3}{\mu^u_0}-
(\lambda\cos\theta_d+\sin\theta_d)\frac{\mu^d_3}{\mu^d_0}e^{-i\phi}
\end{array}\right.\nonumber\\
&&\hspace{4cm}\begin{array}{c}
-\cos\theta_u\sin\theta_de^{i\phi}+\sin\theta_u\cos\theta_de^{-i\phi}\\
\cos\theta_u\cos\theta_d+\sin\theta_u\sin\theta_de^{i2\phi}\\
(-\lambda\sin\theta_de^{2i\phi}+\cos\theta_de^{i\phi})\frac{\mu^u_3}{\mu^u_0}
+(\lambda\sin\theta_d-\cos\theta_d)\frac{\mu^d_3}{\mu^d_0}
\end{array}\nonumber\\
&&\hspace{4cm}\left.\begin{array}{c}
(\lambda\cos\theta_ue^{i\phi}+\sin\theta_u)\frac{\mu^d_3}{\mu^d_0}
-(\lambda\cos\theta_u+\sin\theta_u)\frac{\mu^u_3}{\mu^u_0}e^{-i\phi}\\
(-\lambda\sin\theta_ue^{2i\phi}+\cos\theta_ue^{i\phi})\frac{\mu^d_3}{\mu^d_0}
+(\lambda\sin\theta_u-\cos\theta_u)\frac{\mu^u_3}{\mu^u_0})\\
1
\end{array}\right).
\end{eqnarray}
\par
We examine our model numerically. 
The present experimental values for the quark masses and CKM matrix are given 
in the PDG 2004 \cite{PDG04}; 
\begin{eqnarray}
&&\frac{m_d}{m_s}=0.057\pm0.023,\ \frac{m_s}{m_b}=0.024\pm0.006,\ 
m_b=4.25\pm0.15{\rm GeV},\nonumber\\
&&\frac{m_u}{m_c}=0.0022\pm0.0010,\ \frac{m_c}{m_t}=0.0070\pm0.0007,\ 
m_t=178{ +10.4\atop-8.3}{\rm GeV},\nonumber\\
&&|V_{CKM}|=\left(
\begin{array}{ccc}0.9739 {\rm \ to\ } 0.9751&0.221 {\rm \ to\ } 0.227&0.0029 
{\rm \ to\ }0.0045\\
0.221 {\rm \ to\ } 0.227&0.9730 {\rm \ to\ } 0.9744&0.039 
{\rm \ to\ }0.044\\
0.0048 {\rm \ to\ } 0.014&0.037 {\rm \ to\ } 0.043&0.9990 
{\rm \ to\ }0.9992
\end{array}
\right),\\
&&{\rm vertex\ coordinate\ of\ unitarity\  triangle}\ \ \bar{\rho}=0.20\pm0.09,\ 
\bar{\eta}=0.33\pm0.05,\nonumber\\  
&&{\rm invariant\ measure\ of\ CP\ violation}\ \ J=(2.88\pm0.33)\times10^{-5}. 
\nonumber
\end{eqnarray}
Using the computer simulation, we search the allowed values of 10 parameters 
satisfying the experimental values (24), and then we obtain the following 
values;
\begin{eqnarray}
&&\mu_0^d=4.25\pm0.15{\rm GeV},\ \frac{\mu_1^d}{\mu_0^d}=0.0133\pm0.0027,\ \ 
\frac{\mu_2^d}{\mu_0^d}=-0.0113\pm0.0027,\nonumber\\
&&\hspace{2cm} \frac{\mu_3^d}{\mu_0^d}=0.0260\pm0.0017,\nonumber\\
&&\mu_0^u=178{+10.4\atop-8.3}{\rm GeV},\ \frac{\mu_1^u}{\mu_0^u}=0.00393\pm0.00003,
\ \ \frac{\mu_2^u}{\mu_0^u}=-0.00380\pm0.00003,\\
&&\hspace{2cm} \frac{\mu_3^u}{\mu_0^u}=-0.0150\pm0.0003,\nonumber\\
&&\lambda=0.219\pm0.005,\ \ \phi=-(76.8\pm1.8)^\circ.\nonumber
\end{eqnarray}
Thus, our model can produce all experimental values exactly. 
\par
We consider the meaning of the Cabibbo angle and CP violation phase in our model. 
$|V_{\rm CKM}|_{12}$ elements is expressed approximately in Eq.(23), as 
$|-\cos\theta_u\sin\theta_de^{i\phi}+\sin\theta_u\cos\theta_de^{-i\phi}|$.
This is estimated approximately by the approximate expressions (19) and (21) 
for $\theta_d$ and  $\theta_u$, as 
\begin{eqnarray}
&&[(\cos\theta_u\sin\theta_d-\sin\theta_u\cos\theta_d)^2\cos^2\phi+
(\cos\theta_u\sin\theta_d+\sin\theta_u\cos\theta_d)^2\sin^2\phi]^{1/2}\nonumber\\
&&\approx |\sin(\theta_d+\theta_u)\sin\phi| \approx |\lambda\sin\phi|.\nonumber
\end{eqnarray} 
We can say that the origin of the Cabibbo angle is the ratio $\lambda=\langle|H_D^1
|\rangle_0/\langle|H_D^2|\rangle_0$ 
and the phase of $H_D^1$, and the origin of the CP violation phase is the phase 
of $H_D^1$.
\section{Lepton mass and mixing}
In this section, we consider the lepton mass hierarchy and leptonic mixing 
having the one-maximal and one-large character. We assume the Yukawa interaction 
as Eq. (11) for masses of charged lepton (e, $\mu$, $\tau$) and Dirac masses of 
($\nu_{e}$, $\nu_{\mu}$, $\nu_{\tau}$) and the Yukawa interaction as Eq. (14) 
for Majorana mass; 
\begin{eqnarray}
&&-{\cal L_D}^{l,\nu}=\Gamma_S^{l,\nu}\bar{f^L_S}f^R_SH_S+\Gamma_D^{1l,\nu}
{\bar{\bi f}^L_D}{\bi f}^R_DH_S+\Gamma_D^{2l,\nu}[(\bar{f_1^L}f_2^R+
\bar{f_2^L}f_1^R)H_D^{1}+(\bar{f_1^L}f_1^R-\bar{f_2^L}f_2^R)H_D^2]\nonumber\\
&&\hspace{2cm}+\Gamma_D^{3l,\nu}(\bar{\bi f}_D^L{\bi H}_Df_S^R+
\bar{f_S^L}^t\!{\bi H}_D{\bi f}_D^R)+h.c.,\ \nonumber\\
&&-{\cal L_M}=\Gamma_S^M{^t}\nu^R_SC^{-1}\nu^R_S+\Gamma_D^M{^t}{\bi \nu}^R_D
C^{-1}{\bi \nu}^R_D.\\
&&
\left(\begin{array}{c}{\bi f}_D^{L,R}\\
f_S^{L,R}\end{array}\right)=
\left(\begin{array}{c}f_1^{L,R}\\f_2^{L,R}\\
f_S^{L,R}\end{array}\right)=
\left(\begin{array}{c}{\nu_e}^{L,R}\\{\nu_\mu}^{L,R}\\
{\nu_\tau}^{L,R}\end{array}\right),
\left(\begin{array}{c}e^{L,R}\\ \mu^{L,R}\\
\tau^{L,R}\end{array}\right), \ \ \ 
\left(\begin{array}{c}{\bi H}_D\\H_S\end{array}\right)=
\left(\begin{array}{c}H_D^1\\H_D^2\\H_S\end{array}\right),\nonumber
\end{eqnarray}
where $C$ is the charge conjugation matrix. 
In lepton sector, we do not consider the CP violation, then the phases of 
the fields related are not considered. 
Thus we get the mass matrices for charged lepton ($e$, $\mu$, $\tau$) and 
Dirac mass for neutrino ($\nu_e$, $\nu_\mu$, $\nu_\tau$) and Majorana mass as 
follows; 
\begin{equation}
\begin{array}{l}
M_l=\left(\begin{array}{ccc}
     \mu_1^l+\mu_2^l&\lambda\mu_2^l&\lambda\mu_3^l\\
     \lambda\mu_2^l&\mu_1^l-\mu_2^l&\mu_3^l\\
     \lambda\mu_3^l&\mu_3^l&\mu_0^l
     \end{array}\right),\\ 
M_\nu^D=\left(\begin{array}{ccc}
     \mu_1^{\nu}+\mu_2^{\nu}&\lambda\mu_2^{\nu}&\lambda\mu_3^{\nu}\\
     \lambda\mu_2^{\nu}&\mu_1^{\nu}-\mu_2^{\nu}&\mu_3^{\nu}\\
     \lambda\mu_3^{\nu}&\mu_3^{\nu}&\mu_0^{\nu}
     \end{array}\right),\ \ 
M_M=\left(\begin{array}{ccc}
     M_1&0&0\\0&M_1&0\\0&0&M_0
     \end{array}\right).     
\end{array}     
\end{equation}
Through the see-saw mechanism, neutrino grows up Majorana neutrino and masses 
fall down extremely small because of the GUT scale of Majorana mass, and then 
the mass matrix for left-handed Majorana neutrino is expressed as 
\begin{eqnarray}
&&M_\nu^M=M_\nu^DM_M^{-1}{^t}M_\nu^D=\nonumber\\
&&\left(\begin{array}{ccc}
\frac{(\mu_1^\nu+\mu_2^\nu)^2+\lambda^2{\mu^\nu}_2^2}{M_1}+\frac{\lambda^2
{\mu_3^\nu}^2}{M_0}&
\lambda(\frac{2\mu_1^\nu\mu_2^\nu}{M_1}+\frac{{\mu_3^\nu}^2}{M_0})&
\lambda(\frac{\mu_3^\nu(\mu_1^\nu+\mu_2^\nu)+\mu_2^\nu\mu_3^\nu}{M_1}+
\frac{\mu_0^\nu\mu_3^\nu}{M_0})\\
\lambda(\frac{2\mu_1^\nu\mu_2^\nu}{M_1}+\frac{{\mu_3^\nu}^2}{M_0})&
\frac{(\mu_1^\nu-\mu_2^\nu)^2+\lambda^2{\mu_2^\nu}^2}{M_1}+\frac{{\mu_3^\nu}^2}
{M_0}&
\frac{\lambda^2\mu_2^\nu\mu_3^\nu+\mu_3^\nu(\mu_1^\nu-\mu_2^\nu)}{M_1}+\frac
{\mu_0^\nu\mu_3^\nu}{M_0}\\
\lambda(\frac{\mu_3^\nu(\mu_1^\nu+\mu_2^\nu)+\mu_2^\nu\mu_3^\nu}{M_1}+
\frac{\mu_0^\nu\mu_3^\nu}{M_0})&
\frac{\lambda^2\mu_2^\nu\mu_3^\nu+\mu_3^\nu(\mu_1^\nu-\mu_2^\nu)}{M_1}+
\frac{\mu_0^\nu\mu_3^\nu}{M_0}&
\frac{\lambda^2{\mu_3^\nu}^2+{\mu_3^\nu}^2}{M_1}+\frac{{\mu_0^\nu}^2}{M_0}
\end{array}
\right).
\end{eqnarray}
\par
Diagonalising the mass matrix (27) of charged lepton and (28) of neutrino 
by unitary matrices $U(\mu_i^l)$ and $U(\mu_i^\nu,M_i)$, the MNS leptonic 
mixing matrix $V_{MNS}$ is defined as  
\begin{eqnarray}
&&{\rm diag}(m_e, m_\mu, m_\tau)=U^{-1}(\mu_i^l)M_lU(\mu_i^l),
\ \ \ ^t(e^m, \mu^m, \tau^m)_{L, R}=U^{-1}(\mu_i^l)^t(e, \mu,\tau)_{L, R},
\nonumber\\
&&{\rm diag}(m_{\nu_e}, m_{\nu_\mu}, m_{\mu_\tau})=U^{-1}(\mu_i^\nu,M_i)
M_\nu^MU(\mu_i^\nu, M_i),\\
&&\hspace{2cm} ^t(\nu_e^m, \nu_\mu^m, \nu_\tau^m)_{L}=U^{-1}(\mu_i^\nu, M_i)
^t(\nu_e, \nu_\mu, \nu_\tau)_{L},\nonumber\\
&&V_{MNS}=U(\mu_i^l)^{\dagger}U(\mu_i^\nu,M_i), \nonumber
\end{eqnarray}
where $(e^m, \mu^m, \tau^m)$ and $(\nu_e^m, \nu_\mu^m, \nu_\tau^m)$ are mass 
eigen states of charged lepton $(e, \mu, \tau)$ and Majorana neutrino $(\nu_e, 
\nu_\mu, \nu_\tau)$.
\par
Here, we show the leptonic mixing $V_{\rm MNS}$ has one-maximal and one-large 
mixing character. 
As shown later, $\mu_1^l$, $\mu_2^l$, $\mu_3^l$ are very small compared to 
$\mu_0^l$, then $U(\mu_i^l)$ is nearly equal to identity. 
Thus leptonic mixing is almost equal to neutrino mixing; $V_{\rm MNS}\approx U
(\mu_i^\nu,M_i)$. 
Now, if $M_0\gg M_1$ in Eq. (28), second terms in every elements are 
negligible. 
The parameter values for $\mu_i^\nu$ can be assumed $\mu_1^\nu\approx-
\mu_2^\nu, \mu_3^\nu \ll\mu_0$ and $\lambda\approx 0.22$ as estimated in 
previous quark sector, then, if we set $\mu_1^\nu-\mu_2^\nu\approx\mu_3^\nu$, 
$\mu_1^\nu+\mu_2^\nu=\delta\ll\mu_3^\nu$ and neglect the 
$\lambda^2$ term, we can get the following expression
\begin{equation}
M_\nu^M\approx
\left(\begin{array}{ccc}
\frac{{\delta}^2}{M_1}&-\lambda\frac{{\mu_3^\nu}^2}{2M_1}&
-\lambda\frac{{\mu_3^\nu}^2-3\delta\mu_3^\nu}{2M_1}\\
-\lambda\frac{{\mu_3^\nu}^2}{2M_1}&
\frac{{\mu_3^\nu}^2}{M_1}&
\frac{{\mu_3^\nu}^2}{M_1}\\
-\lambda\frac{{\mu_3^\nu}^2-3\delta\mu_3^\nu}{2M_1}&
\frac{{\mu_3^\nu}^2}{M_1}&
\frac{{\mu_3^\nu}^2}{M_1}
\end{array}
\right).
\end{equation} 
This matrix can be diagonalized by the unitary matrix 
$U(\mu_i^\nu,M_i)$ as 
\begin{eqnarray}
&&{\rm diag}\left(
\frac{\mu_3^2}{M_1}\left\{
\frac12\left[\frac1{\sqrt{1+(\frac{3\lambda}{\sqrt{2}}\frac{\mu_3}{\delta})^2}}
+1\right](\frac{\delta}{\mu_3})^2-\frac{(\frac{3\lambda}{2})^2}
{\sqrt{1+(\frac{3\lambda}{\sqrt{2}}\frac{\mu_3}{\delta})^2}}
\right\}, \right.\nonumber\\
&&\hspace{1cm}\left.\frac{\mu_3^2}{M_1}\left\{
\frac12\left[-\frac1{\sqrt{1+(\frac{3\lambda}{\sqrt{2}}\frac{\mu_3}
{\delta})^2}}+1\right](\frac{\delta}{\mu_3})^2+\frac{(\frac{3\lambda}{2})^2}
{\sqrt{1+(\frac{3\lambda}{\sqrt{2}}\frac{\mu_3}{\delta})^2}}
\right\},
 \frac{2{\mu_3^\nu}^2}{M_1}
\right),\nonumber\\
&&\approx U(\mu_i^\nu,M_i)^{-1}M_\nu^MU(\mu_i^\nu,M_i),\\
&&U(\mu_i^\nu,M_i)\approx
\left(\begin{array}{ccc}
1&0&0\\
0&1/\sqrt{2}&1/\sqrt{2}\\
0&-1/\sqrt{2}&1/\sqrt{2}
\end{array}\right)
\left(\begin{array}{ccc}
1&0&\eta\\
0&1&0\\
-\eta&0&1
\end{array}\right)
\left(\begin{array}{ccc}
\cos\theta&\sin\theta&0\\
-\sin\theta&\cos\theta&0\\
0&0&1
\end{array}\right),\\
&&\hspace{1cm}\eta\approx-\frac{\lambda}{2\sqrt{2}}(1-\frac{\delta}{2\mu_3}),
\ \ \ \ \tan2\theta\approx\frac{3\lambda}{\sqrt{2}}\frac{\mu_3}{\delta}.
\nonumber
\end{eqnarray}
From this approximate expression, it is recognized that $\nu_\mu\mbox{-}\nu_\tau$ 
mixing becomes maximal and $\nu_e\mbox{-}\nu_\mu$ mixing angle can be large, 
for example, if $\delta\approx\lambda\mu_3$, $\dis{\tan2\theta\approx
\frac{3}{\sqrt{2}}}$. Furthermore, it is recognized that $|V_{\rm MNS}|_{13}\approx
\eta\approx\lambda/2\sqrt{2}. 
$
\par
Next, we examine our model numerically. Firstly, we analyze the charged lepton 
sector; estimate $\mu_i^l$ and $U(\mu_i^l)$ satisfying experimental data using 
Eqs. (27) and (29).
Using the mass of ($e$, $\mu$, $\tau$);
\begin{equation}
\frac{m_e}{m_\mu}=0.004835\pm0.000005,\ \frac{m_\mu}{m_\tau}=0.05946\pm0.00001,\ 
m_\tau=1776.99{+0.29\atop-0.26}{\rm MeV},
\end{equation}  
we can get the values of parameters and charged lepton mixing matrix; 
\begin{equation}
\begin{array}{l}
\mu_0=1777{\rm MeV},\ \ \lambda=0.219\pm0.005\ (\rm in\mbox{-}put\ determined \ 
from\ quark\ sector\ analysis),\\
\displaystyle{\frac{\mu_1^l}{\mu_0^l}}=0.03007\pm0.00005,\ \ 
\displaystyle{\frac{\mu_2^l}{\mu_0^l}}=-(0.02900\pm0.00005),\ \ 
\displaystyle{\frac{\mu_3^l}{\mu_0^l}}=0.0250\pm0.0017,\\
U(\mu_i^l)=\left(
\begin{array}{ccc}
0.993\sim0.994&-(0.113\sim0.114)&0.005\sim0.006\\
0.113\sim0.114&0.993&0.025\sim0.028\\
-(0.008\sim0.009)&-(0.024\sim0.027)&1.00
\end{array}\right).
\end{array}
\end{equation}
Present experimental status of neutrino mixing is summarized as \cite{ATMOS,
SOLAR,CHOOZ},
\begin{eqnarray}
&&\sin^22\theta_{\rm atm}>0.9,\ \ 1.3\times10^{-3}{\rm eV}^2<\Delta m^2_{\rm atm}
<3.0\times10^{-3}{\rm eV}^2, \nonumber\\
&&3.0\times10^{-1}<\tan^2\theta_{\odot}<5.8\times10^{-1},\nonumber\\
&&\hspace{4cm}5.9\times10^{-5}{\rm eV}^2<\Delta m^2_{\odot}<9.3\times10^{-5}
{\rm eV}^2,\ \ \\
&&{[V_{\rm MNS}]_{13}}^2<0.067. \nonumber
\end{eqnarray}
We assume that $\dis{\Delta m^2_{\rm atm}=m_{\nu_\tau}^2-m_{\nu_\mu}^2}$, 
$\dis{\Delta m^2_{\odot}=m_{\nu_\mu}^2-m_{\nu_e}^2}$, $\dis{\theta_{\rm atm}=
\nu_\mu\mbox{-}\nu_\tau\ {\rm mixing\ angle}}$, $\dis{\theta_{\odot}=
\nu_e\mbox{-}\nu_\mu\ {\rm mixing\ angle}}$, then the ratios of neutrino masses and 
magnitudes of elements of $V_{\rm MNS}$ are restricted as 
\begin{eqnarray}
&&0<\frac{m_{\nu_e}}{m_{\nu_\mu}}<0.6,\ \ 0.14<\frac{m_{\nu_\mu}}{m_{\nu_\tau}}<
0.24,\ \ 0.036{\rm eV}< m_{\nu_\tau}<0.055{\rm eV}\nonumber\\
&&|V_{\rm MNS}|=\left(
\begin{array}{ccc}
|c_{\odot}c_{13}|&|s_{\odot}c_{13}|&|s_{13}|\\
|-s_{\odot}c_{\rm atm}-c_{\odot}s_{\rm atm}s_{13}|&|c_{\odot}c_{\rm atm}-s_{\odot}
s_{23}s_{13}|&|s_{\rm atm}c_{13}|\\
|s_{\odot}s_{\rm atm}-c_{\odot}c_{\rm atm}s_{13}|&|-c_{\odot}s_{\rm atm}-s_{\odot}
c_{\rm atm}s_{13}|&|c_{\rm atm}c_{13}|
\end{array}\right)\nonumber\\
&&\hspace{1.3cm}=\left(\begin{array}{ccc}
0.77\ {\sim}\ 0.88&0.46\ {\sim}\ 0.61&0.00\ {\sim}\ 0.26\\
0.10\ {\sim}\ 0.49&0.47\ {\sim}\ 0.78&0.57\ {\sim}\ 0.81\\
0.28\ {\sim}\ 0.61&0.34\ {\sim}\ 0.71&0.57\ {\sim}\ 0.81
\end{array}\right),\\
&&{\rm where\ }c_{\odot}=\cos\theta_{\odot}, \ \ s_{\odot}=\sin\theta_{\odot},\ \ 
c_{\rm atm}=\cos\theta_{\rm atm},\ \ s_{\rm atm}=\sin\theta_{\rm atm}.
\nonumber
\end{eqnarray}
Using the Eqs. (27), (28), (29) and the numerical result (34) for charged lepton 
and experimental data (36) for neutrino, we estimate the allowed values for 
parameters $\mu_i^\nu$; 
\begin{eqnarray}
&&\lambda=0.219\pm0.005\ (\rm in\mbox{-}put\ determined \ 
from\ quark\ sector\ analysis),\nonumber\\
&&\frac{\mu^{\nu}_1}{\mu^{0}}=0.050\pm0.007,\ \
\frac{\mu^{\nu}_2}{\mu^{0}}=-(0.021\pm0.007),\ \
\frac{\mu^{\nu}_3}{\mu^{0}}=0.052\pm0.010,\\ 
&&\frac{M_1}{M_0}=0.0020\pm0.0007.\nonumber
\end{eqnarray}
For these allowed parameters, the mass of $\nu_e$ and $\dis{|V_{\rm MNS}|_{13}}$ 
are rather restricted as 
\begin{eqnarray}
&&\frac{m_{\nu_e}}{m_{\nu_\mu}}=0.36\sim0.49,\ \
|V_{\rm MNS}|_{13}=0.04\sim0.06.
\end{eqnarray}
\section{Conclusion}
We assumed the weak bases of flavors $\dis{(u,c)_{L,R},\ (d,s)_{L,R},\ 
(e,\mu)_{L,R}}$ and Dirac neutrino $\dis{(\nu_e,\nu_\mu)_{L,R}}$ are the $S_3$ 
doublet and $\dis{t_{L,R}, b_{L,R}, \tau_{L,R}, {\nu_\tau}_{L,R}}$ are the $S_3$ 
singlets. 
Further, we assumed the Higgs $S_3$ doublet $\dis{(H_D^1,H_D^2)}$ and 
Higgs $S_3$ singlet $H_S$. In general, though $\dis{H_D^1, H_D^2, H_S}$ are 
complex, $\dis{H_D^2, H_S}$ can be made real by the $SU(2)_L$ gauge freedom of 
$\dis{(u,c)_{L},\ (d,s)_{L},\ (e,\mu)_{L},\ (\nu_e,\nu_\mu)_{L}}$.  
From these $S_3$ doublets and singlets, we constructed  $S_3$ invariant Yukawa 
interactions and hermitian mass matrices for weak basis of flavor. 
In our model, because the way to construct an $S_3$ invariant Yukawa interaction 
is unique, we can define the weak basis of flavor unambiguously.  
\par
Obtained mass 
matrices for quark sector are 
\begin{equation}
M_d=\left(\begin{array}{ccc}
     \mu_1^d+\mu_2^d&\lambda\mu_2^de^{i\phi}&\lambda\mu_3^d
     e^{i\phi}\\
     \lambda\mu_2^de^{-i\phi}&\mu_1^d-\mu_2^d&\mu_3^d\\
     \lambda\mu_3^de^{-i\phi}&\mu_3^d&\mu_0^d
     \end{array}\right),\ \ 
M_u=\left(\begin{array}{ccc}
     \mu_1^u+\mu_2^u&\lambda\mu_2^ue^{-i\phi}&\lambda\mu_3^u
     e^{-i\phi}\\
     \lambda\mu_2^ue^{i\phi}&\mu_1^u-\mu_2^u&\mu_3^u\\
     \lambda\mu_3^ue^{i\phi}&\mu_3^u&\mu_0^u
     \end{array}\right), \nonumber
\end{equation}
where $\dis{\lambda=\langle|H_D^1|\rangle_{0}|/\langle|H_D^2|\rangle_{0}}$ and 
$\dis{\phi={\rm phase\ of\ }\langle H_D^1\rangle_{0}}$.
From the present experimental data for quark masses and $\dis{V_{\rm CKM}}$ matrix 
involving the CP violation \cite{PDG04}, we can get the results; 
\begin{eqnarray}
&&\frac{\mu_1^d}{\mu_0^d}=0.0133\pm0.0027,\ \frac{\mu_2^d}{\mu_0^d}=-0.0113\pm
0.0027,\nonumber\ \frac{\mu_3^d}{\mu_0^d}=0.0260\pm0.0017,\nonumber\\
&&\frac{\mu_1^u}{\mu_0^u}=0.00393\pm0.00003,
\ \ \frac{\mu_2^u}{\mu_0^u}=-0.00380\pm0.00003,\ \ 
\frac{\mu_3^u}{\mu_0^u}=-0.0150\pm0.0003,\nonumber\\
&&\lambda=0.219\pm0.005,\ \ \phi=-(76.8\pm1.8)^\circ.\nonumber
\end{eqnarray}
In our model, the origin of the Cabibbo angle is the ratio $\dis{\lambda=
\langle|H_D^1|\rangle_{0}/\langle|H_D^2|\rangle_{0}}$ and the origin of the CP 
violation is the phase of $H_1$.
\par
For lepton sector, mass matrices are obtained as 
\begin{equation}
M_l, M_\nu^D=\left(\begin{array}{ccc}
     \mu_1^{l,\nu}+\mu_2^{l,\nu}&\lambda\mu_2^{l,\nu}&\lambda\mu_3^{l,\nu}\\
     \lambda\mu_2^{l,\nu}&\mu_1^{l,\nu}-\mu_2^{l,\nu}&\mu_3^{l,\nu}\\
     \lambda\mu_3^{l,\nu}&\mu_3^{l,\nu}&\mu_0^{l,\nu}
     \end{array}\right),\ \ 
M_M=\left(\begin{array}{ccc}
     M_1&0&0\\0&M_1&0\\0&0&M_0
     \end{array}\right). \nonumber    
\end{equation}
In our model, one-maximal and one-large mixing angle character of the lepton 
mixing matrix $\dis{V_{\rm MNS}}$ is obtained naturally from the hierarchy of mass 
parameters  $\dis{\mu_1^\nu+\mu_2^\nu}$ $\dis{\ll\mu_1^\nu, -\mu_2^\nu, \mu_3^\nu\ll
\mu_0^\nu}$, $\dis{M_1\ll M_0}$ and smallness of $\dis{\lambda\sim0.22}$ 
without any other symmetry restriction. 
From the present experimental data for charged lepton mass \cite{PDG04} and 
neutrino mass and mixing $\dis{V_{\rm MNS}}$ \cite{ATMOS,SOLAR,CHOOZ}, 
we obtained the allowed values for mass parameters; 
\begin{eqnarray}
&&\displaystyle{\frac{\mu_1^l}{\mu_0^l}}=0.03007\pm0.00005,\ \ 
\displaystyle{\frac{\mu_2^l}{\mu_0^l}}=-(0.02900\pm0.00005),\ \ 
\displaystyle{\frac{\mu_3^l}{\mu_0^l}}=0.0250\pm0.0017,\nonumber\\
&&\frac{\mu^{\nu}_1}{\mu^{0}}=0.050\pm0.007,\ \
\frac{\mu^{\nu}_2}{\mu^{0}}=-(0.021\pm0.007),\ \
\frac{\mu^{\nu}_3}{\mu^{0}}=0.052\pm0.010,\nonumber\\ 
&&\frac{M_1}{M_0}=0.0020\pm0.0007.\nonumber
\end{eqnarray}
In this allowed values, the $\nu_e$ mass are estimated rather large as 
$\dis{\frac{m_{\nu_e}}{m_{\nu_\mu}}=0.36\sim0.49}$ and $\dis{|V_{\rm MNS}|_{13}}$ 
is restricted as $\dis{0.04\sim0.06}$.


\begin{thebibliography}{999}
\bibitem{ATMOS}  
 K. Nishikawa, presented at the XI Int. Symp. on Lepton and Photon Interactions 
 at High Energies(Lepton Photon 2003), 
 Fermilab, August, 2003
\bibitem{SOLAR}
Y. Kosio, to appear in the Proceedings of 38th Rencontres de Moriond on Electroweak 
Interactions and Unified Theories, Les Arcs, France, March 15-22, 2003, 
hep-ex/0306002. \\
The SNO Collaboration (S. Ahmed {\it et al.}), nucl-ex/0309004.\\
The KamLAND Collaboration (K. Eguchi {\it et al.}), Phys. Rev. Lett. {\bf 90},
 021802(2003). 
\bibitem{HARARI} 
 H. Harari, H. Haut and J. Weyers, Phys. Lett. {\bf B78}(1978),459.
\bibitem{KOIDE1}
 Y. Koide, Phys. Rev. {\bf D28}(1983),252.
\bibitem{TANIMOTO1}
 M. Tanimoto, Phys. Rev. {\bf D41}(1990),1586.
\bibitem{KOIDE2} 
 Y. Koide and H. Fusaoka, Z. Phys. {\bf C71} (1996), 459. 
\bibitem{GATTO} 
 R. Gatto, G. Marchio, G. Sartori and F. Strocchi, Nucl. Phys. {\bf B163}(1980), 
 221.
\bibitem{BRANCO1}
 G. C. Branco, J. I. Silva-Marcos and M. N. Rebelo, Phys. Lett. {\bf B237}(1990), 
 446.
\bibitem{TESHIMA1}
 T. Teshima and T. Sakai,  Prog. Thor. Phys. 97(1997), 653. 
\bibitem{TESHIMA2}
 T. Teshima and T. Asai,  Prog. Thor. Phys. 105(1997), 763. 
\bibitem{TESHIMA3}
 T. Teshima, T. Asai and Y. Abe, Phys. Rev. {\bf D66}(2002),093011.
\bibitem{BRANCO2} 
 E.~Kh.~Akhmedov, G.~C.~Branco, F.~R.~Joaquim and J.~I.~Silva-Marcos, 
    Phys. Lett. {\bf B498}(2001), 237.
\bibitem{BRANVCO3} 
 G.~C.~Branco and J.~I.~Silva-Marcos, Phys.~Lett.~{\bf B526}(2002),104.
\bibitem{FRITZSCH}  
 H. Fritzsch, Phys. Lett. {\bf B73}(1978),317; Nucl Phys. {\bf B155}(1979),189.
\bibitem{FUKUGITA1}
 M. Fukugita, M. Tanimoto, T. Yanagida, Prog. Theor. Phys. {\bf 89}(1993), 263.  
\bibitem{FUKUGITA2}
 M. Fukugita, M. Tanimoto, T. Yanagida, Phys. Lett. {\bf B562}(2003), 273. 
\bibitem{PAKUBASA}
 S. Pakvasa and H. Sugawara, Phys. Lett. {\bf 73B}(1978), 61.\\
 S. Pakvasa and H. Sugawara, Phys. Lett. {\bf 82B}(1979), 105.
\bibitem{SUGAWARA}
 Y. Yamanaka, H. Sugawara and S. Pakvasa, Phys. Rev. {\bf D25}, (1982)1895.
\bibitem{KUBO}
 J. Kubo, A. Mondrag\'on, M. Mondrag\'on and E. Rodr\'iguez-Jauregui, Prog. Theor. 
 Phys. {\bf 109}(2003) 795. 
\bibitem{SEESAW} T.~Yanagida, in {\it Proceedings of the Workshop on the 
    Unified Theories and Baryon Number in the Universe} Tsukuba, {\it 1979}, 
    ed. O.~Sawada and A. Sugamoto, KEK report No.79-18, Tsukuba (1979), p.~95;
    \ 
    M.~Gell-Mann, P. Ramond and R. Slansky, in {\it Supergravity, Proceedings 
    of the Workshop, Stony Brook, New York, 1979}, ed. P.~van Nieuwenhuizen 
    and D.~Freedmann (North-Holland, Amsterdam, 1979), p.~315.
\bibitem{PDG04} S. Eidelman {\it et. al.}( Particle Data Group), Phys. Lett. 
{\bf B592}(2004), 1. 
\bibitem{CHOOZ}
The CHOOZ Collaboration (M. Apollonio {\it et al.}), Eur. Phys. J. {\bf C27},
331(2003).   
\end{thebibliography}
\end{document}